\documentclass{INTERSPEECH2023}

\interspeechcameraready

\usepackage{amsmath}
\usepackage{multirow}
\usepackage{amssymb}
\usepackage{comment}
\usepackage{tablefootnote}
\usepackage{subfig}
\usepackage{color}
\usepackage{threeparttable}

\newlength\savedwidth
\newcommand{\wcline}[1]{\noalign{\global\savedwidth\arrayrulewidth\global\arrayrulewidth 1.0pt} \cline{#1}
\noalign{\global\arrayrulewidth\savedwidth}}
\usepackage{bm}

\newcommand{\norm}[1]{\left\lVert#1\right\rVert}

\makeatletter
\newcommand{\figcaption}[1]{\def\@captype{figure}\caption{#1}}
\newcommand{\tblcaption}[1]{\def\@captype{table}\caption{#1}}
\makeatother

\title{CAPTDURE: Captioned Sound Dataset of Single Sources}
\name{Yuki Okamoto$^1$, Kanta Shimonishi$^1$, Keisuke Imoto$^2$,\\ Kota Dohi$^3$, Shota Horiguchi$^3$, Yohei Kawaguchi$^3$}
\address{
  $^1$Ritsumeikan University, Japan\\
  $^2$Doshisha University, Japan \\
  $^3$Hitachi, Ltd., Japan}
\email{y-okamoto@ieee.org, is0460rf@ed.ritsumei.ac.jp, keisuke.imoto@ieee.org, \{kota.dohi.gr,shota.horiguchi.wk,yohei.kawaguchi.xk\}@hitachi.com}

\begin{document}

\maketitle

\begin{abstract}
%\vspace{-2pt}
%Methods of environmental sound separation and synthesis using language queries such as captions for sound have been proposed.
%In conventional studies on environmental sound separation and synthesis using captions, captions for multiple-source sounds, which included multiple sound-event sources, were used.
%In conventional studies on environmental sound separation and synthesis using captions, captions for multiple-source sounds were used.
In conventional studies on environmental sound separation and synthesis using captions, datasets consisting of multiple-source sounds with their captions were used for model training.
However, when we collect the captions for multiple-source sound, it is not easy to collect detailed captions for each sound source, such as the number of sound occurrences and timbre.
Therefore, it is difficult to extract only the single-source target sound by the model-training method using a conventional captioned sound dataset.
In this work, we constructed a dataset with captions for a single-source sound named CAPTDURE, which can be used in various tasks such as environmental sound separation and synthesis.
Our dataset consists of 1,044 sounds and 4,902 captions.
We evaluated the performance of environmental sound extraction using our dataset.
The experimental results show that the captions for single-source sounds are effective in extracting only the single-source target sound from the mixture sound.
\end{abstract}


%\vspace{-2pt}
\section{Introduction}
%\vspace{-2pt}
Environmental sounds are indispensable to enhance the sense of presence and immersion in media content such as movies and games.
One way to prepare a desired single-source sound is to obtain it from a sound database \cite{Gemmeke_ICASSP_2017}. 
However, the required environmental sounds may not always exist in the database.
%There are many environmental sounds on the Internet, but they are multiple-source sounds from multiple sound events, and it is not easy to obtain only the single-source target sound.
Moreover, there are many multiple-source sounds on the Internet, but obtaining only the single-source target sound is not easy.

To obtain only the single-source target sound, environmental sound extraction and separation methods have been proposed \cite{Sudo_IROS_2019,ochiai_INTERSPEECH_2020,Kavalerov_WASPPA_2019,Delcroix_TASLP_2022,Lee_ISMIR_2019,Tzinis_ICASSP_2020} for extracting only the single-source target sound from a multiple-source sound.
For example, a method has been proposed with which a sound event class is specified and only sounds corresponding to the sound event class are extracted \cite{Sudo_IROS_2019,ochiai_INTERSPEECH_2020,Delcroix_TASLP_2022,Slizovskaia_ICASSP_2109}.
To extract sounds more accurately than the sound event class, the method using onomatopoeic words that transcribes the characteristics of the sound to be extracted has been proposed \cite{Okamoto_ICASSP_2022}.
However, since onomatopoeic words are language and culture dependent, the dataset for model training needs to be modified in accordance with the user's native language.
To avoid dependence on language and culture, a method of environmental sound extraction and separation using captions for sound has also been proposed \cite{Liu_INTERSPEECH_2022,Kilgour_INTERSPEECH_2022,Dong_ICLR_2023}.
With this method, the characteristics of the multiple-source target sound are expressed by captions and the multiple-source target sound can be extracted.
%This method enables us to extract only the target sound corresponding to the caption in which characteristics of the target sound.
This method enables us to extract the multiple-source target sound, which includes some sound event classes corresponding to the caption.
The conventional environmental sound-extraction method using captions is based on a dataset \cite{kim_NAACL_2019,Drossos_ICASSP_2020,Wu_2019_ICASSP} in which captions are assigned to multiple-source sounds.
The conventional dataset 
Table \ref{table:examples_captions} shows captions collected for single- and multiple-source sounds in our dataset. 
As shown in Table \ref{table:examples_captions}, captions collected for multiple-source sounds are less detailed for a single sound event than those collected for a single-source sound.
Therefore, it is difficult to extract only a single-source target sound using captioned sound dataset collected for multiple-source sounds.
We need the caption for that single-source sound to extract a single-source target sound from multiple-source sounds. 

%Another method to obtain the required sound, environmental sound synthesis methods, which artificially generate target sounds using , have also been proposed \cite{Kong_ICASSP2019,okamoto_ATSIP_2022,Liu_MLSP_2021,Chen_TIP_2020,Zhou_CVPR_2018}.
%Another method to obtain the required sound, environmental sound synthesis methods, which artificially generate target sounds using some input information, such as image, sound event class, and natural language, have also been proposed \cite{Kong_ICASSP2019,okamoto_ATSIP_2022,Liu_MLSP_2021,Chen_TIP_2020,Zhou_CVPR_2018}.
%One of the techniques of environmental sound synthesis using captions have been proposed \cite{Kreuk_ICLR_2023,Liu_arXiv_2023,Yang_arXiv_2022,Huang_arXiv_2023}.
Environmental sound synthesis methods, which artificially generate sounds, have also been proposed to obtain the required sound \cite{Kong_ICASSP2019,okamoto_ATSIP_2022,Liu_MLSP_2021,Chen_TIP_2020,Zhou_CVPR_2018}. 
Recently, there are some studies of environmental sound synthesis using captions \cite{Kreuk_ICLR_2023,Liu_arXiv_2023,Yang_arXiv_2022,Huang_arXiv_2023}.
%Another method to obtain the required sound, the techniques of environmental sound synthesis using captions, have also been proposed \cite{Kreuk_ICLR_2023,Liu_arXiv_2023,Yang_arXiv_2022,Huang_arXiv_2023}.
However, no captioned sound datasets represent only a single-source sound, so it is difficult to fine-tune the generated single-source sound.
We need the caption corresponding to a single-source sound to control synthesized environmental sound finely.

\begin{table}[t!]
%\vspace{-5pt}
\caption{Examples of captions in CAPTDURE. The sound sources are \underline{underlined}.}
\vspace{-8pt}
\label{table:examples_captions}
\centering
\resizebox{\linewidth}{!}{%
    \begin{tabular}{@{}lp{80mm}@{}}
       \toprule
       \#Sources&Caption\\\midrule
        Single & \begin{minipage}[t]{\linewidth}\begin{itemize}
            \item One or two \underline{keyboards} continue to be pressed slowly with a light, high-touch sound.
            \item The sound of the buttons on the \underline{keyboard} being pressed one by one quite slowly.
            %\item Slowly and firmly tap the \underline{keys}, which are quite heavy like typewriters.
            \item The electronic tone of the digital alarm \underline{clock} is high-pitched and continues to sound slowly to gradually faster.
            \item The bright, high-pitched alarm \underline{clock} bells are ringing.
        \end{itemize}\end{minipage}\\\midrule
        Multiple &  \begin{minipage}[t]{\linewidth}\begin{itemize}
            \item \underline{Keyboard} typing and \underline{mouse} clicking noises are continually heard.
            \item The sound of a \underline{toaster} rang out as if to counteract the sound of typing on a small \underline{keyboard}.
            \item The alarm \underline{clock} is drowned out by the loud noise of the hair \underline{dryer}.
            \item The alarm \underline{clock} beeps at equal intervals, with one final sound to close the \underline{lock}.
            %\item \underline{Keyboards} pound rhythmically and hair dryers make noisy noises.
        \end{itemize}\end{minipage}\\
        \bottomrule
    \end{tabular}%
}
\vspace{-10pt}
\end{table}
\begin{figure*}[t!]
\begin{minipage}[c]{.60\linewidth}
%\vspace{-5pt}
\tblcaption{List of recording equipment}
\vspace{-8pt}
\label{table:record_equipment}
\centering
\resizebox{\linewidth}{!}{%
\begin{tabular}{@{}llll@{}}
    \wcline{1-4}
     & \\[-8pt]
    {\bf Recording equipment} & {\bf Software} & {\bf Microphone} & {\bf Audio interface}\\
    \hline \hline
    & \\[-8pt]
    Apple/Macbook Pro (16-inch 2019) & Audacity\tablefootnote{\url{https://www.audacityteam.org/}} & SHURE/MX150B/O-XLR & Roland/Rubix24 \\
    TASCAM/DR-44WL & Built-in & Built-in & Built-in\\
    Apple/iPhone SE & PCM recording\tablefootnote{\url{https://ko-yasui.com/}} & Built-in & Built-in\\
    Huawei/dtab d01-G & AudioRec\tablefootnote{\url{https://audiorec.jp.aptoide.com/app}} & Built-in & Built-in \\
    \wcline{1-4}
\end{tabular}
}
\vspace{6pt}
%\end{minipage}
    %\begin{minipage}[c]{.60\linewidth}
        \tblcaption{Recorded sound event classes}
        \setlength{\tabcolsep}{3pt}
        \vspace{-8pt}
        \label{table:sound_event_list_tmp}
        \centering
        \resizebox{\linewidth}{!}{%
        \begin{tabular}{@{}llcrr@{}}
            \wcline{1-5}
             &\\[-8pt]
             {\bf Sound event class} &  {\bf Description} & {\bf \#Subclasses} &{\bf \#Clips} & {\bf Duration [s]} \\
            \hline \hline
            &\\[-8pt]
             Keyboard & Sound of keyboard typing & 5 & 60 & 465.0\\
             Door & Sound of door opening and closing & 4 & 48 & 279.0\\
             Mouse & Sound of mouse clicking  & 7 & 84 & 600.0\\
             Water tap running & Sound of water tap running & 5 & 60 & 431.0\\
             Dryer & Operating sound of dryer & 6 & 252 & 1,746.0\\
             Ventilation fan & Operating sound of ventilation fan & 3 & 36 & 284.0\\
             Door lock & Sound of door locking & 6 & 48 & 245.0\\
             Intercom & Sound of intercom & 5 & 56 & 350.0\\
             Door knock & Sound of door knock & 6 & 72 & 430.0\\
             Microwave & Operating sound of microwave & 4 & 48 & 329.0\\
             Toaster & Operating sound of toaster & 5 & 60 & 465.0\\
             Cutlery & Sound of hitting cutlery & 7 & 52 & 362.0\\
             Clock & Operating sound of alarm clock & 5 & 60 & 420.0 \\
             Fan & Operating sound of fan & 3 & 108 & 779.0 \\
             \cline{1-5}
             &\\[-8pt]
             Total & & 71 & 1,044 & 7,185.0\\
            \wcline{1-5}
        \end{tabular}
        }
    \end{minipage}
    \hfill
    \begin{minipage}[c]{.36\linewidth}
        \centering
        \subfloat[Single-source sounds\label{fig:hist_single_tmp}]{%
        \includegraphics[width=\linewidth]{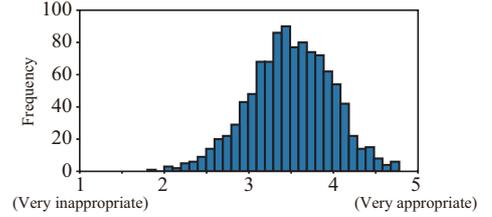}%
        }\\
        \vspace{4pt}
        \subfloat[Multiple-source sounds\label{fig:hist_mixture_tmp}]{%
        \includegraphics[width=\linewidth]{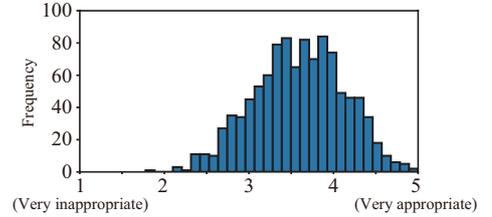}%
        }
        \caption{Histograms of average appropriateness score}
        \label{fig:hist}
    \end{minipage}
\vspace{-8pt}
\end{figure*}
%
%In this paper, we construct a dataset with sound descriptions for single sound sources that can be used in various tasks that utilize environmental sounds. 
In this paper, we constructed a dataset with captions for sound shown in Table \ref{table:examples_captions} for single sound sources that can be used in various tasks that involve environmental sounds.
The dataset is called CAPTDURE (CAPTioned sound Dataset of single soURcEs), pronounced as ``capture.''
As an example of the use of the dataset, we conduct an experiment on environmental sound extraction using captions.
In our experiment, we show that the use of captions for sound assigned to a single-source sound is effective in extracting only target sounds in a multiple-source sound.

The rest of the paper is organized as follows.
In Sec.~\ref{construct_dataset}, we describe the creation of CAPTDURE.
In Sec.~\ref{evaluation}, we discuss our experiments on environmental sound extraction using language query. 
Finally, we summarize and conclude this paper in Sec.~\ref{conclusion}.
%

%\vspace{-2pt}
\section{Creation of CAPTDURE}
%\vspace{-2pt}
\label{construct_dataset}

\subsection{Design of CAPTDURE}
%\vspace{-2pt}
The CAPTDURE dataset consists of the following contents.
\begin{itemize}
  \setlength{\itemsep}{0pt}
  \item{\bf Single-source sounds}\\
   We recorded a total of 1,044 single-source sounds, consisting of  14 types of daily sound events, as shown in Table \ref{table:sound_event_list_tmp}.
   For each sound event, we recorded approximately 40 to 100 sounds while changing the recording conditions, such as recording equipment.
   The length of each sound was set to 5 to 9 seconds.
   Each sound event class was divided into subclasses in accordance with differences in timbre and pitch.
  \item{\bf Multiple-source sounds}\\
  We created a total of 1,044 multiple-source sounds using the recorded single-source sounds.
  We selected two sounds of different sound event classes from the recorded sounds and mixed the sounds so that the signal-to-noise ratio (SNR) was 0 dB.
  We padded with zero the shorter sound and aligned the sound length of the two sounds.
  %\item{\bf Captions for each single- and multiple-source sounds}
  %\begin{itemize}
  \item{\bf Captions for single-source sounds}\\
  We collected a total of 4,902 captions (3 or more captions per single-source sound).
  We describe the details of the captions collection in Sec.\ref{collectibg_caption}.
  \item{\bf Captions for multiple-source sounds}\\ 
  We collected a total of 3,132 captions (3 captions per multiple-source sound).
  %\end{itemize}
  \item{\bf Appropriateness score for each caption}\\
  We asked crowdworkers to score the appropriateness level for captions transcribed by others.
  The appropriateness scores enable us to evaluate captions on the basis of the judgment of others.
  We describe the details of the appropriateness scores in Sec.~\ref{caption_evaluation}.
  \item{\bf Worker ID}\\
  CAPTDURE includes anonymized IDs of workers who gave captions and appropriateness scores.
\end{itemize}
This dataset is freely available online\tablefootnote{\url{https://sites.google.com/view/yuki-okamoto/dataset}}.
\subsection{Recording environment and setup}
%\vspace{-2pt}
\label{recordin_environment}
The environmental sounds included in CAPTDURE were recorded in two types of soundproof rooms.
The first soundproof room is a length of 3.1 m, a width of 5.4 m, and a height of 2.7 m with a reverberation time ($T_{60}$) of 0.2 s.
The second soundproof room is a length of 4.9 m, a width of 4.0 m, and a height of 2.5 m with a reverberation time ($T_{60}$) of 0.2 s.
Some environmental sounds, such as ``water tap running'' and ``door'' that are difficult to record in a soundproof room, were recorded at locations other than these rooms. 
Table \ref{table:record_equipment} lists the equipment used for the sound recording. 
The sampling frequency was 48 kHz, and the quantization bit rate was 16 bits. 
The microphone was set at a distance of approximately 0.3 to 0.5 m from the sound source being recorded. 
%We recorded when the sound volume of the source was extremely loud.
%The distance between the object and the microphone was kept at sufficient distance.

%\subsection{Recorded sound event class}
%\label{recording_sound_event}
 %We recorded the sounds of 14 types of daily sound events, as shown in Table \ref{table:sound_event_list_tmp}.
%For each sound event, we recorded approximately 40 to 100 sounds while changing the recording conditions, such as recording equipment.
%The length of each sound was set to 5 to 9 seconds.
%Each sound event class was divided into subclasses according to differences in the way the sound was played and the type of sound source.
%Each sound-event class was divided into subclasses in accordance with differences in timbre and pitch.

\begin{comment}
%
\begin{figure*}[t!]
\centering
\resizebox{\linewidth}{!}{%
\includegraphics[scale=1.0]{figures/CaptionSeparation_ConvTasNet}%
}
\caption{Architecture of Conv-TasNet based environmental-sound-extraction method using caption.}
\label{fig:propsed_method}
\end{figure*}
%
\end{comment}

\subsection{Captions collection}
%\vspace{-2pt}
\label{collectibg_caption}
%singel source caption
%ja average words: 22.85089399744572
%en average words: 10.522030651340996
%mixture caption
%ja average words: 36.65357598978289
%en average words: 16.96264367816092
%In this paper, we used Lancers, which is a crowdsourcing service\tablefootnote{\url{https://www.lancers.jp/}} to collect sound captions from Japanese crowd workers.
We used Lancers\tablefootnote{\url{https://www.lancers.jp/}}, which is a crowdsourcing service in Japan, to collect captions for sounds from Japanese crowdworkers.
In the pre-experiment, we tried using Amazon Mechanical Turk\tablefootnote{\url{https://www.mturk.com/}} (MTurk).
However, comparing Lancers and MTurk, we confirmed that the quality of crowdworkers was higher in Lancers.

The collection of captions was conducted by presenting five sounds to one worker.
In the pre-experiment, the five sounds presented to the crowdworkers were from different sound event classes.
%In this case, we confirmed that the detail of the caption given to each sound, such as "the sound of the microwave," decreased.
However, the caption of each sound mainly consisted of the caption of the sound class it belongs to.
Therefore, the five sounds presented to the crowdworkers consisted of the same-sound event class, and we instructed the crowdworkers to write a different sentence as the caption given to each sound.
Crowdworkers were also instructed to avoid providing only with the type of sound, such as ``the sound of the microwave,'' or only with onomatopoeic words, such as /k a N k a N/.
 
We collected a total of 4,902 captions (3--5 captions per single-source sound) for the single-source sounds.
Moreover, we collected a total of 3,132 captions (3 captions per multiple-source sound) for the multiple-source sounds.
In addition to the captions collected in Japanese, CAPTDURE also contains captions translated into English using the DeepL API\tablefootnote{\url{https://www.deepl.com/en/docs-api}}.
%Table \ref{table:examples_captions} shoes examples of the collected captions written in Japanese converted to the English using DeepL API.

\subsection{Captions evaluation}
%\vspace{-2pt}
\label{caption_evaluation}
We asked crowdworkers to score the level of appropriateness for captions transcribed by others. 
We present pairs of audio samples and the corresponding caption to a crowdworker. 
The crowdworker then gives an appropriateness score for each caption.
%The appropriate score is on a scale of five from 1 (not very appropriate) to 5 (very appropriate) for each caption by others, who were not the worker giving captions.
The appropriateness score is on a five-point scale, 1 (very inappropriate) to 5 (very appropriate), for each caption by those who did not create the caption.
The appropriateness scores for each caption were collected from more than three crowdworkers.

Figure \ref{fig:hist} shows the histogram of each collected appropriateness score calculated as the average per sound for single and multiple-source sounds.
Most of the captions were given an appropriateness score of 3 or higher.
This fact indicates that we were able to collect many appropriate captions to express each sound.
%Thus, we were able to collect many appropriate captions to express each sound.

\subsection{Data splitting}
%\vspace{-2pt}
\label{data_split}
%現在追記中です。
We split the single- and multiple-source sounds of each subclass as approximately 7:1:2 for the train, validation, and test sets.
The captions corresponding to each sound are also included in each set.
The statistics of each set are shown in Table \ref{table:statitics_information}.
%The number of words in captions in each set was divided so that the number of words in train, validation, and test was similar.
We artificially split the sounds of each set to balance the number of words/characters of captions.

%We also split multiple-source sounds as well as single sounds.
%To avoid any unbalance in the subclasses contained in the multiple-source sound, we divided the sound into train, validation, and test at a ratio of approximately 7:1:2.
%The number of words in captions in each dataset was also divided so that the number of words in train, validation, and test was similar.

%We split recorded sounds and collected captions, as shown in Table \ref{table:statitics_information}.

%
%
\begin{table}[t!]
%\vspace{-5pt}
\caption{Statistics of train, validation, and test set of CAPTDURE}
\vspace{-8pt}
\label{table:statitics_information}
\centering
\resizebox{\linewidth}{!}{%
\begin{tabular}{@{}lrrc@{}}
    \wcline{1-4}
     & \\[-8pt]
    {\bf Dataset} & {\bf \#Clips} & {\bf \#Captions} & {\bf \#Words (en) / \#Characters (ja)}\\
    \hline \hline
    & \\[-8pt]
    {\bf Single-source sound} & & &\\
    \quad train & 795 & 3,774 & 10.53 (en) / 22.80 (ja) \\
    \quad validation & 82 & 374 & 10.62 (en) / 22.91 (ja)\\
    \quad test & 167 & 501 & 10.43 (en) / 23.05 (ja) \\
    \cline{1-4}
    &\\[-8pt]
    {\bf Multiple-source sound} & & &\\
    \quad train & 795 & 2,385 & 16.83 (en) / 36.47 (ja) \\
    \quad validation & 82 & 246 & 16.60 (en) / 35.53 (ja)\\
    \quad test & 167 & 754 & 17.79 (en) / 38.06 (ja) \\
    \wcline{1-4}
\end{tabular}%
}
%\vspace{-10pt}
\end{table}
%
%\vspace{-2pt}
\section{Experiments}
%\vspace{-2pt}
\label{evaluation}
To demonstrate the efficiency of CAPTDURE, we evaluated the performance of environmental sound extraction using caption.

\begin{comment}
%
\begin{figure*}[t!]
\centering
\resizebox{\linewidth}{!}{%
\includegraphics[scale=1.0]{figures/CaptionSeparation.eps}%
}
\caption{Architecture of U-Net based environmental-sound-extraction method using caption.}
\label{fig:propsed_method}
\end{figure*}
%
\end{comment}

\subsection{Network architecture}
%\vspace{-2pt}
%Figure \ref{fig:propsed_method} shows the network architecture of using our experiment.
%Following the conventional sound extraction method, we use the Conv-TasNet \cite{Luo_TASLP_2019} based sound separation architecture.
%We trained the Conv-TasNet \cite{Luo_TASLP_2019}-based model to extract the sound that corresponds to caption ${\bm l}$ from mixture ${\bm x}$.
To extract the sound that corresponds to caption ${\bm l}$ from multiple-source sound ${\bm x}$, we trained a neural network based on Conv-TasNet \cite{Luo_TASLP_2019}.
%We use the Conv-TasNet \cite{Luo_TASLP_2019} based sound separation architecture.
The $T$-length multiple-source sound $\bm{x}\in\mathbb{R}^{1 \times T}$ is fed to the encoder, which consists of a one-dimensional (1-D) convolution layer as follows:
\begin{equation}
{\bm W}=\mathsf{Encoder}\left(\bm{x}\right)\in\mathbb{R}^{C\times T'}.
\end{equation}

At the same time, the sound caption ${\bm l}$ is fed to the Bidirectional Encoder Representations from Transformers (BERT) \cite{Devlin_NAACL_2019}.
In this experiment, we used the pre-trained English\tablefootnote{\url{https://huggingface.co/bert-base-uncased}} and Japanese\tablefootnote{\url{https://huggingface.co/cl-tohoku/bert-base-japanese}} BERT models.
The vector obtained by BERT is compressed by linear transformation $\mathsf{Linear}(\cdot)$ to the $C$-dimension as follows:
\begin{equation}
{\bm o}=\mathsf{Linear}\left(\mathsf{BERT}\left(\bm{l}\right)\right)\in\mathbb{R}^{C}.
\end{equation}
%
%The obtained embedding ${\bm o}$ is stretched in the time directions to form $T'$ feature maps $\bmit{O} = \left[\bmit{o}_1,\dots,{\bmit{o}}_{T'}\right]$ where ${\bmit{o}}_{t'}\in\mathbb{R}^{1\times C}$ for $t'\in\left\{1,\dots,T'\right\}$ is the matrix the in which all elements are ${\bmit{o}}_{t'}$.
%The obtained matrix $w$ and $o$ are computed the Hadamard product $\odot$ and fed to $\mathsf{Separator}(\cdot)$, which consist of $K$-stacked 1-D convolution layers as follows:
The soft-mask $\bm{M}$ is calculated from the obtained matrix $\bm{W}$ and vector $\bm{o}$ as
\begin{equation}
%{\bm{z}}&=\bm{w}\odot\bm{o}\in\mathbb{R}^{T'\times C},\\
\bm{M}=\mathsf{Separator}\left(\vphantom{\bm{W}\odot\left[\vphantom{\bm{o},\bm{o},\dots,\bm{o}}\right.}\right.
\bm{W}\odot[\underbrace{\bm{o},\bm{o},\dots,\bm{o}}_{T'}]
\left.\vphantom{\bm{W}\odot\left.\vphantom{\bm{o},\bm{o},\dots,\bm{o}}\right]}\right)\in\left(0,1\right)^{C\times T'},
\end{equation}
where $\odot$ denotes the Hadamard product and $\mathsf{Separator}(\cdot)$ is $K$-stacked 1-D convolutional layers.
Finally, the Hadamard product of soft-mask ${\bm M}$ and ${\bm W}$ is fed to the decoder, which consists of a 1-D convolution layer, to obtain the target signal corresponding to the caption as follows:
\begin{equation}
\hat{{\bm y}}=\mathsf{Decoder}\left(\bm{M}\odot\bm{W}\right)\in\mathbb{R}^{1 \times T}.
\end{equation}

The network was trained to minimize the L1 norm between target sound ${\bm y}$ and estimated sound $\hat{{\bm y}}$.
%During model training, the loss function is defined as the L1 norm between target sound ${\bm y}$ and estimated sound $\hat{{\bm y}}$.

%
%
\begin{table}[t!]
%\vspace{-5pt}
\small
\caption{Experimental conditions}
\vspace{-8pt}
\label{table:experiment}
\centering
%\resizebox{\linewidth}{!}{%
\begin{tabular}{@{}ll@{}}
    \wcline{1-2}
     &\\[-8pt]
    {\bf Waveform} & \\
    \quad Multiple-source-sound length & \SI{10}{\second} \\
    \quad Sampling rate & \SI{16}{kHz}\\
    \quad Waveform encoding & 16-bit linear PCM \\
    \cline{1-2}
    &\\[-8pt]
    {\bf Parameters of Conv-TasNet} & \\
    \quad \# of filters in autoencoder & 256 \\ 
    \quad Length of filters in samples  & 20 \\
    \quad \# of channels in bottleneck & 256 \\ 
    \quad \# of channels in convolutional block & 512 \\
    \quad Kernel size in convolutional block & 3 \\
    \quad \# of convolutional blocks in each repeat & 8 \\
    \quad \# of repeats & 4 \\
    \cline{1-2}
    &\\[-8pt]
    {\bf Parameters of model training} & \\
    \quad Batch size & 1 \\
    \quad Training epoch & 120 \\
    \quad Learning rate & 0.0001 \\
    \quad Optimizer  & RAdam \cite{RAdam_ICLR_2020} \\
    \wcline{1-2}
\end{tabular}
%}
\vspace{-10pt}
\end{table}
\begin{table*}[t!]
%\vspace{-5pt}
\small
\caption{SDRi [dB] for extracted signals}
\vspace{-8pt}
\label{table:result}
\centering
\resizebox{\linewidth}{!}{%
\begin{tabular}{@{}llccccc@{}}
    \wcline{1-7}
    & \\[-8pt]
    & &\multicolumn{5}{c}{SNR} \\
     \cline{3-7}\\[-8pt]
    Dataset & Model-training method&\SI{-10}{\dB} & \SI{-5}{\dB} & \SI{0}{\dB} & \SI{5}{\dB} & \SI{10}{\dB}\\
    \cline{1-7}
    & \\[-8pt]
    \multirow{4}{*}{Inter-event-class dataset} & Training using multiple-source caption (ja) &$5.12 \pm 4.22$ & $4.39 \pm 4.04$ & $3.00 \pm 4.21$ & $0.93 \pm 4.92$ & $-1.85 \pm 5.95$ \\
    & Training using multiple-source caption (en) &$4.67 \pm 4.03$ & $3.88 \pm 3.88$ & $2.43 \pm 4.04$ & $0.17 \pm 4.71$ & $-2.78 \pm 5.67$ \\
    & {\bf Training using single-source caption (ja)} &$7.28 \pm 5.19$ & $6.26 \pm 5.02$ & $4.63 \pm 5.18$ & $2.09 \pm 5.71$ & $-1.08 \pm 6.54$ \\
    & {\bf Training using single-source caption (en)} &$6.42 \pm 4.84$ & $5.14 \pm 4.66$ & $3.13 \pm 4.89$ & $0.40 \pm 5.54$ & $-2.91 \pm 6.40$ \\
    \cline{1-7}
    & \\[-8pt]
    \multirow{4}{*}{Intra-event-class dataset} & Training using multiple-source caption (ja) &$3.22 \pm 2.78$ & $2.49 \pm 2.37$ & $1.11 \pm 2.31$ & $-1.01 \pm 3.08$ & $-3.93 \pm 4.34$ \\
     & Training using multiple-source caption (en) &$3.09 \pm 2.50$ & $2.38 \pm 2.11$ & $1.15 \pm 2.09$ & $-0.86 \pm 2.90$ & $-3.61 \pm 4.22$ \\
    & {\bf Training using single-source caption (ja)} &$4.36 \pm 3.60$ & $3.24 \pm 3.18$ & $1.36 \pm 3.01$ & $-1.33 \pm 3.66$ & $-4.69 \pm 4.73$ \\
    & {\bf Training using single-source caption (en)} &$4.47 \pm 3.09$ & $3.23 \pm 2.41$ & $1.29 \pm 2.49$ & $-1.48 \pm 3.27$ & $-4.87 \pm 4.42$ \\
    \wcline{1-7}
\end{tabular}%
}
\end{table*}
\begin{figure*}[t!]
\vspace{-3pt}
\centering
\includegraphics[scale=0.85]{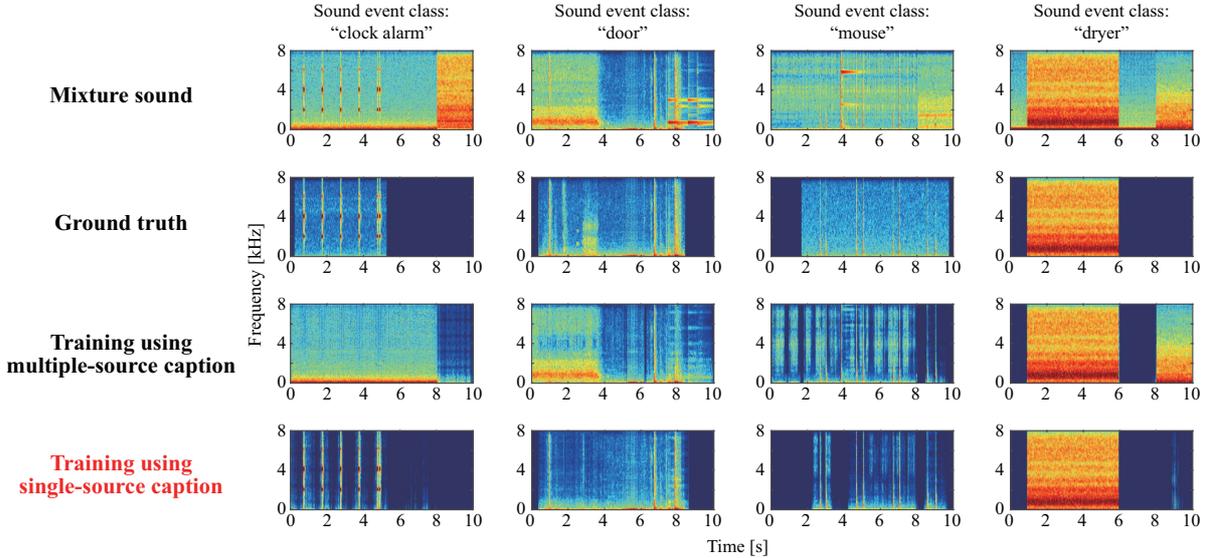}
\vspace{-6pt}
\caption{Examples of environmental sound extraction using captions for the inter-event-class dataset.  Mixture spectrogram (first row), ground truth spectrogram (second row), results of model training using multiple-source caption (third row), and results of model training using single-source caption (proposed) (fourth row).}
\label{fig:spectro_result}
\vspace{-6pt}
\end{figure*}
\subsection{Dataset construction}
%\vspace{-2pt}
\label{dataset_construction}
We randomly selected three single sounds with different sound event classes from CAPTDURE to create a mixture sound.
%The sounds of each sound event class were divided at a ratio of approximately 7:2:1 and used as training, validation, and evaluation data, respectively.
The training and validation sets consisted of 2,385 and 246 mixture sounds, respectively.
We used three randomly selected captions for each sound for our experiments.

We constructed two evaluation datasets using the test set of single-source sounds from CAPTDURE: the inter-event-class dataset and the intra-event-class dataset. Each sound in those datasets is a mixture sound of a target sound and an interference sound, but those are from different sound event classes in the inter-event-class dataset and from the same sound event class in the intra-event-class dataset.
The signal-to-interference ratio was varied by $\{-10, -5, 0, 5, 10\}$ \si{\dB}.
Each evaluation set consisted of 501 mixture sounds for each SNR. 

\vspace{-2pt}
\subsection{Training and evaluation setup}
%\vspace{-2pt}
Table \ref{table:experiment} lists the experimental conditions and parameters used for this experiment.
In this experiment, we compared two types of model training methods as follows:
\begin{itemize}
\item{\bf Training using multiple-source captions}\\
   This model training method uses the captions corresponding to the multiple-source sounds.
   This method is equivalent to the model training method using the conventional dataset.
   \\[-6pt]
  \item{\bf Training using single-source captions}\\
  This model training method uses captions corresponding to single-source sounds.
\\[-8pt]
\end{itemize}
In the evaluation, we evaluate the sound extraction performance using the evaluation dataset created in Sec.~\ref{dataset_construction} for each model training method.

To evaluate each model-training method, we used signal-to-distortion ratio improvement (SDRi) \cite{SDR_TASLP_2006} as an evaluation metric, which is defined as the difference between the SDR of the target sound to the mixture sound and that of the target sound to the extracted sound.
\begin{comment}
\begin{equation}
    \mbox{SDRi} = 10\log_{10}\left(\frac{\norm{\bm{y}}^2}{\norm{\bm{y} - \hat{\bm{y}}}^2}\right) - 10\log_{10}\left(\frac{\norm{\bm{y}}^2}{\norm{\bm{y} - \bm{x}}^2}\right).
\end{equation}
\end{comment}
%We conducted evaluations regarding SDRi on each evaluation dataset in Sec.~\ref{dataset_construction}.
We evaluated in terms of the SDRi on each evaluation dataset.

\subsection{Experimental results}
%\vspace{-2pt}
Table \ref{table:result} lists the SDRi on each evaluation dataset.
The training using single-source captions enables us to extract only the target sound from a mixture sound compared with the training using multiple-source captions. 
Moreover, from the results of the intra-event-class dataset, the caption for a single-source sound is effective to extract only the target sound, even when the interference sound from the same sound event class appears in the mixture sound.

Figure \ref{fig:spectro_result} shows the spectrogram of the extracted sounds using single-source captions.
We used four audio samples in the inter-event-class dataset with \SI{0}{\dB}.
We observed that the training using multiple-source captions left a significant amount of non-target sounds.
%We observed that the sounds extracted with the model trained using multiple-source captions contain a significant amount of non-target sounds.
The training using single-source captions extracted only the single-source target sound.
On the other hand, the model trained using single-source captions could extract only the single-source target sound.
Thus, the captions for a single-source source are effective in extracting only the single-source target sound.  
%
%
\begin{comment}
\begin{table*}[t!]
\vspace{-5pt}
\small
\caption{SDRi [dB] for extracted signals}
\vspace{2pt}
\label{table:result}
\centering
\resizebox{\linewidth}{!}{%
\begin{tabular}{@{}llccccc@{}}
    \wcline{1-7}
     & \\[-8pt]
    & & \multicolumn{5}{c}{SNR} \\
     \cline{3-7}\\[-8pt]
    Network &Model training method&\SI{-10}{\dB} & \SI{-5}{\dB} & \SI{0}{\dB} & \SI{5}{\dB} & \SI{10}{\dB}\\
    \cline{1-7}
    & \\[-8pt]
    \multirow{4}{*}{U-Net} & Training by mixture source caption (en) &$ \pm $ & $ \pm $ & $ \pm $ & $ \pm $ & $ \pm $ \\
    & Training by mixture source caption (jp) &$ \pm $ & $ \pm $ & zs$ \pm $ & $ \pm $ & $ \pm $ \\
    &{\bf Training by single source caption (en)} &$ \pm $ & $ \pm $ & $ \pm $ & $ \pm $ & $ \pm $ \\
    &{\bf Training by single source caption (jp)} & $ \pm $ & $ \pm $ & $ \pm $ & $ \pm $ & $ \pm $ \\
    \cline{1-7}
    & \\[-8pt]
     \multirow{4}{*}{Conv-TasNet} & Training by mixture source caption (en) &$ \pm $ & $ \pm $ & $ \pm $ & $ \pm $ & $ \pm $ \\
    & Training by mixture source caption (jp) &$4.76 \pm 3.83$ & $3.85 \pm 3.74$ & $2.28 \pm 3.87$ & $-0.07 \pm 4.48$ & $-3.16 \pm 5.42$ \\
    & {\bf Training by single source caption (en)} &$ \pm $ & $ \pm $ & $ \pm $ & $ \pm $ & $ \pm $ \\
    & {\bf Training by single source caption (jp)} &$6.84 \pm 4.43$ & $5.78 \pm 4.28$ & $3.97 \pm 4.47$ & $1.35 \pm 5.07$ & $-2.03 \pm 5.91$ \\
    \wcline{1-7}
\end{tabular}%
}
\end{table*}
%
\end{comment}

%\vspace{-2pt}
\section{Conclusion}
%\vspace{-2pt}
\label{conclusion}
We constructed a dataset with captions for a single-source sound that can be used in various tasks that use environmental sounds.
We recorded a total of 1,044 single-source sounds and collected a total of 4,902 captions for recorded single-sound sources.
The experimental results indicate that the use of sound captions assigned to a single-sound source is effective in extracting only the target sounds in mixture sounds.
Verifying our dataset's effectiveness for other tasks involving environmental sounds is necessary. 

%\lipsum[66]

%\section{Acknowledgements}

\newpage
\bibliographystyle{IEEEtran}
\bibliography{mybib}

\end{document}